\documentclass[preprint,prl,aps]{revtex4}

\newcommand{\be}{\begin{equation}}
\newcommand{\ee}{\end{equation}}
\newcommand{\ba}{\begin{eqnarray}}
\newcommand{\ea}{\end{eqnarray}}
\newcommand{\ii}{\'{\i}}

\newcommand{\nn}{\nonumber}
\newcommand{\expq}{e_q}
\newcommand{\lnq}{\ln_q}

\newcommand{\tr}{{\mathrm{Tr}}}

\draft

\begin{document}
\title{$q$-Thermodynamics: First law for
quasi-stationary states}
\author{S. Mart\'{\i}nez${^{1,\,2}}$\thanks{%
E-mail: martinez@venus.fisica.unlp.edu.ar} and A.
Plastino${^{1,\,2}}$%
\thanks{
E-mail: plastino@venus.fisica.unlp.edu.ar}.}
\address{$1$ Instituto de F\'{\i}sica de La Plata,
National University La Plata, C.C.%
727 , 1900 La Plata, Argentina. \\ $^2$ Argentine
National
Research Council (CONICET)}


\begin{abstract}

We discuss peculiar aspects of the first law of
thermodynamics for
systems characterized by the presence of
meta-equilibrium
quasi-stationary states for which the pertinent
phase/configuration spaces is generally {\it
inhomogeneous}. As a
consequence, the naive additivity requirement for
thermodynamic
quantities ceases to be satisfied.\vspace{0.2 cm}


\vspace{1 cm}
\end{abstract}

\maketitle

\newpage
\section{Introduction}

The requirement of additivity for certain
thermodynamic quantities places
strict constraints with regards to the symmetries of
the concomitant phase (or
configuration) space and is indivisibly linked with
the {\it homogeneity} of
the system under consideration, an assumption that
remains  frequently
unmentioned (possibly because it is often fulfilled).
{\it Today}, exotic and
complex thermodynamic systems or processes are the
subject of considerable
attraction: colossal magneto-resistance manganites,
amorphous and glassy
nano-clusters, high-energy collision processes, etc.,
characterized by the
common feature of non-equilibrium states stationary
for significantly long
periods of time (compared to typical time-scales of
their microscopic
dynamics). Scale invariance and hierarchical
structures are here preserved, but
the pertinent phase/configuration spaces are generally
{\it inhomogeneous}. As
a consequence, the naive additivity requirement ceases
to be satisfied.

The existence of $N$-body systems characterized by the
presence of
meta-equilibrium quasi-stationary states (QSS) has
been conclusively proven and
in these cases traditional thermostatistics displays
some shortcomings. The
best theoretical description that has been thus far
obtained uses the
strictures of non-extensive thermostatistics (NET)
\cite{rapisarda}.
 Non-extensive
thermostatistics is by now considered as a new
paradigm for statistical
mechanics~\cite{tsallisURL,t_csf6,t_bjp29,libro,pp_bjp29,pennini}.
It is based on Tsallis' non-extensive information
measure~\cite{t_jsp52}
\begin{equation}
S_q = k_B\,\frac{1-\sum p_n^{\ q}}{q-1},
\end{equation}
where $k_B$ stands for Boltzmann constant, to be set
equal to
unity herefrom, and $\{p_n\}$ is a set of normalized
probabilities. The real parameter $q$ is called the
index of
non-extensivity, the conventional Boltzmann--Gibbs
statistics
being recovered in the limit $q=1$.

We will show in the present effort that {\it for these systems},
and for other that are also amenable to a NET description, the
First Law of Thermodynamics {\it retains its standard form, even
if the pertinent state is not one of standard thermodynamic
equilibrium}. We will also try to provide some insights in what
refers to the peculiar way NET describes thermodynamic   systems.
Such peculiarity partly explains some unfamiliar NET
characteristics.

\section{The NET-normalization problem: TMP vs. OLM}

NET-theory comes in several flavors. The literature on
Tsallis'
thermostatistics considers three possible choices for
the
evaluation of expectation values within the
non-extensive
scenario. As some of the (non-extensive) expectation
values are
always regarded as constraints in the associated
$q$-MaxEnt
approach~\cite{pp_pla177}, three different
NET-probability
distributions will ensue. For the sake of
completeness, a brief
account is given in the Appendix. We will employ here
just one of
them, usually called the
Tsallis--Mendes--Plastino~(TMP)
\cite{TMP} choice, that is today the one preferred by
most NET
researchers. We use it, however, in the guise of what
has been
called \cite{OLM} the ``optimal Lagrange multipliers
(OLM)
approach".

\subsection{TMP expectation values}

If we deal with $W$ microstates and our a priori
knowledge is that
of $M$ expectation values $\langle O_j \rangle=o_j$
(plus
normalization), the quantity to be extremized in order
to obtain
the probability distribution $\{ p_n \}$ that
describes our system
according to Jaynes' MaxEnt procedure
reads~\cite{TMP,jaynes,katz}
\be
F=S_{q}[\{p_n\}] -\lambda_0^{\rm (TMP)}\left( \sum_{n=1}^W
p_n-1\right) -\sum_{j=1}^M\lambda_j^{\rm (TMP)} \left(
\frac{\sum_{n=1}^W p_n^{\ q} \, {o_j}_{\, n}} {\sum_{n'=1}^{W}
p_{n'}^{\ q}}-\langle O_{j}\rangle_{q}\right), \label{F} \ee where
$M+1$ Lagrange multipliers $\lambda_j^{\rm (TMP)}$ have been
introduced (a classical language is being used for the time being
for simplicity's sake). As a result of the MaxEnt variational
procedure \cite{pp_pla177,jaynes,katz} one finds that the Tsallis'
probability distribution has the form
\begin{equation}
p_{n}=\frac{f_{n}^{\ 1/(1-q)}}{\bar{Z}_{q}},
\label{pv}
\end{equation}
where
\begin{equation}
f_{n}= 1-\frac{(1-q)\,\sum_{j=1}^M \lambda_j^{\rm
(TMP)}\left({o_{j}}_{\, n}- \langle
O_{j}\rangle_{q}\right)
}{\sum_{n'=1}^W p_{n'}^{\ q}} \equiv f_n^{\rm (TMP)},
\label{fin}
\end{equation}
is called the configurational characteristic (here the
TMP one)
~\cite{TMP}, that should be positive in order to
guarantee that
the probabilities $p_n$ be real for arbitrary $q$
(Tsallis' cutoff
condition~\cite{pp_pla177,pp_pla193}). The denominator
in
Eq.~(\ref{pv}) (related to the multiplier
$\lambda_0^{\rm (TMP)}$)
is given by
\begin{equation}
\bar{Z}_{q}=\sum_{n} f_{n}^{\ 1/(1-q)}, \label{ZZq}
\end{equation}
and represents a ``pseudo" partition function that in
the $q \rightarrow 1$
limit {\it does not yield the conventional partition
function} $Z_1$ but, instead,
$Z_1 \, \exp{\left(\sum_{j=1}^M \lambda_j \, \langle
O_{j}\rangle\right)}$. Let
us remark that, because of Tsallis' cutoff
\cite{pp_pla193}, the sum over
states $n$ is restricted to those for which $f_n$ is
positive, since otherwise
the condition implies $f_n \equiv 0$.

Notice also that, from Eqs.~(\ref{pv})--(\ref{ZZq}),
 the TMP expression obtained for $p_{n}$
 is {\it
explicitly self-referential}. It is important to
stress that this fact often
leads to numerical difficulties in concrete
applications (see, for instance,
Ref.~\cite{disisto}). Indeed,  it obscures the
underlying physics, because the
concomitant Lagrange multipliers lose their
traditional physical
meaning~\cite{casas}.  This fact led credence to the
belief that classical
thermodynamics is recovered {\it only} in the $q
\rightarrow 1$
limit~\cite{TMP}.

\subsection{The OLM treatment}

In order to overcome the problems mentioned in the last paragraph,
Mart\'{\i}nez {\it et al.} \cite{OLM} devised a method that
straightforwardly avoids the self referential nature of the TMP
probabilities. In the process they discredited  the notion that
classical thermodynamics is recovered {\it only} in the $q
\rightarrow 1$.

The central idea of \cite{OLM} is the introduction of new,
putatively {\it optimal} Lagrange multipliers (OLM) for the
Tsallis' variational problem. Thus, one  is extremize the
$q$-entropy with centered mean values (a legitimate procedure)
which entails recasting the constraints in the fashion

\be \sum_{n=1}^W p_n^{\ q}\left({o_j}_{\, n}-\langle
O_j\rangle_q\right) = 0 \qquad \qquad j=1,\ldots,M,
\label{vm3OLM}
\ee so that one deals now with

\be F=S_{q}[\{p_n\}] -\lambda_0 \left( \sum_{n=1}^{W}
p_{n}-1\right) -\sum_{j=1}^{M}\lambda_j \,
\sum_{n=1}^{W} p_{n}^{\
q}\left( {o_j}_{\, n}-\langle O_{j}\rangle_{q}\right).
\ee The
ensuing microscopic probabilities are, formally, still
given by
Eqs.~(\ref{pv}) and~(\ref{ZZq}), but Eq.~(\ref{fin})
is replaced
by
\begin{equation}
f_{n}=1-(1-q)\sum_{j=1}^M \lambda_j\, \left( {o_j}_{\,
n}-\langle
O_{j}\rangle_{q}\right) \equiv f_n^{\rm (OLM)}.
\label{finueva}
\end{equation}
In this way, the configurational characteristic in OLM
form {\it does not
depend explicitly on the set of probabilities}
$\{p_n\}$. It is  obvious that
the solution of a constrained extremizing problem via
the celebrated Lagrange
method depends exclusively on i) the functional form
one is dealing with and
ii) the constraints. From a mathematical point of
view, the Lagrange
multipliers are just auxiliary quantities to be
eliminated at the end of the
process. As a consequence, TMP and OLM probabilities
should coincide. However,
from a physical point of view the Lagrange Multipliers
are connected with the
{\em intensive} variables of the problem.
For two subsystems in thermodynamic equilibrium the pertinent intensive variables are equal.
Thus, the Lagrange multipliers  are
 important quantities and  one should expect
differences in a system's
description as ``seen" from either the TMP or the OLM
vantage points. Of
course, there exists a straightforward mapping between
the two descriptions
\cite{OLM}. However,  the {\it handling} or {\it
manipulation} is, in the OLM
instance, considerably simpler. Notice that the OLM
variational procedure
solves {\it directly} for the optimized Lagrange
multipliers. Comparing  the
TMP and OLM approaches one realizes that the
concomitant probabilities are
identical if \be \lambda_j = \frac{\lambda_j^{\rm
(TMP)}}{\sum_{n=1}^W p_n^{\
q}} = \bar{Z}_q^{\ q-1} \ \lambda_j^{\rm (TMP)} \qquad
\qquad j=1,\ldots,M,
\label{olvida} \ee where use has been made of the
relation $\sum_n p_n^{\ q}=
\bar{Z}_{q}^{\ 1-q}$~\cite{TMP,OLM} under the
assumption that the available a
priori data is the same for both approaches. Notice
that the two associated
pseudo partition functions (if adequately expressed),
do coincide, being of the
form $\bar Z_q=\{[1+(1-q)\lambda_0]/q\}^{1/(1-q)}$,
with
$\lambda_0=\lambda_0^{\rm (TMP)}$.

The OLM treatment is completed with the definition of
the ``true" (not the
pseudo)  partition function, that does indeed go over
to $Z_1$ in the limit $q
\rightarrow 1$, namely~\cite{OLM},
\begin{equation}
\ln Z_q \equiv \ln \bar{Z}_q - \sum_{j=1}^M \lambda_j
\, \langle
O_j \rangle_q.
\label{lnqz}
\end{equation}
It is important to stress that here, however, the
corresponding
TMP function~\cite{TMP} uses the so-called
$q$-logarithms, $\ln_q
x\equiv(1-x^{1-q})/(q-1)$, instead of the ordinary
ones.

Now, from Eq.~(\ref{lnqz}) one is straightforwardly
led to an important result
~\cite{ley0}
\begin{eqnarray}
\frac{\partial\ }{\partial \langle O_{j}\rangle_q}
\left(\ln{\bar{Z}_{q}}\right) & = & \lambda_j
\label{termo1}
\\
\frac{\partial\ }{\partial \lambda_j}
\left(\ln{Z_{q}}\right) & =
& - \, \langle O_j\rangle_q, \label{termo2}
\end{eqnarray}
for $j=1,\ldots,M$. These equations constitute the
basic information-theory
relations in Jaynes' version of statistical
mechanics~\cite{jaynes,katz}.
Again, notice here the presence of ordinary logarithms
in the OLM instance.
Instead, the TMP formulation has to do with
generalized $q$-logarithms.  Finally, let us remark that the several OLM
applications thus
far developed allow one to appreciate the fact that,
unless two-body
interactions are involved, the results of classical
problems of statistical
mechanics are independent of the
$q$-value~\cite{casas}. Obviously, the OLM
results can easily be translated into TMP language
making use of
Eq.~(\ref{olvida}).

\subsection{The OLM procedure in quantum language}
\label{SECTolm}

 It is convenient now to base the following
considerations on a quantum framework. In  such an
 environment, the main tool is the density
operator
$\hat{\rho}$, that can be obtained by recourse to the
MaxEnt
Lagrange multipliers' method \cite{katz}. Within the
nonextensive
framework one has to extremize the information
measure~\cite{t_jsp52}
\begin{equation}
S_{q}
[\hat\rho]=\frac{1-\tr\,(\hat{\rho}^{\,q})}{q-1},
\label{entropy}
\end{equation}
subject to i) the normalization requirement and ii)
the assumed a
priori knowledge of the generalized expectation values
of, say
$M$, relevant observables, namely
\begin{equation}
\langle\hat{O}_j \rangle_q = \frac{\tr\,(\hat
\rho^{\,q} \,
\hat{O}_j)}{\tr\,({\hat \rho}^{\,q})} \qquad \qquad
j=1,\ldots,M.
\label{gener}
\end{equation}
It is important to recall that, from an Information
Theory view-point,
 equilibrium ensues when these $M$ operators commute
with the Hamiltonian
 \cite{katz}. We do not make such an assumption  here.

 \vskip 3mm
 The quantum  constraints are recast in the following
manner
\begin{eqnarray}
\tr\,(\hat{\rho}) & = & 1, \label{normalization}
\\
\tr\left[ \hat{\rho}^{\,q}\left( \hat{O}_j -
\langle\hat{O}_j\rangle_q\right)
\right] & = & 0 \qquad \qquad j=1,\ldots,M,
\label{constraints}
\end{eqnarray}
where the $q$-expectation values $\{\langle\hat
O_1\rangle_q,\ldots,\langle\hat
O_M\rangle_q\}$ constitute the input a priori
information. Performing the
constrained extremization of Tsallis entropy one
obtains~\cite{OLM}
\begin{equation}
\hat{\rho} = \frac{{\hat{f}_q}^{\;
1/(1-q)}}{\bar{Z}_q}, \label{rho}
\end{equation}
where, if $\{\lambda_1,\ldots,\lambda_M\}$ are the optimal
Lagrange multipliers, and we  define for brevity's sake the
generalized deviations \be \delta_q\hat{O}\equiv
\hat{O}-\langle\hat{O}\rangle_q, \ee then the quantal
configurational characteristic has the form \be \hat{f}_q =
\hat{\openone} - (1-q) \sum_{j=1}^M \, \lambda_j \;
\delta_q\hat{O}_j, \label{charact} \ee if the quantity in the
right-hand side of (\ref{charact}) is positive definite, and
otherwise $\hat f_q=0$ (cutoff condition~\cite{pp_pla193,TMP}).
The normalizing factor in Eq.~(\ref{rho}) corresponds to the OLM
generalized partition function which is given, in analogy with the
classical situation,
by~\cite{OLM,virial,gasideal,temperature,ley0,cuerponegro}

\be \bar{Z}_q = \tr \, \left( {\hat{f}_q}^{\;
1/(1-q)}\right) = \tr \, \left[
\, \expq \left( - \sum_{j=1}^M
\lambda_j~\delta_q\hat{O}_j \right) \right],
\label{Zqbar} \ee where the trace evaluation is to be
performed with due
caution in order to account for the Tsallis  cutoff
and \be \expq (x) \equiv
[1+(1-q)x]^{1/(1-q)}, \label{expq} \ee is a
generalization of the exponential
function, that is recovered when $q\rightarrow 1$.

It is to be pointed out that within the TMP framework
one obtains
from the normalization condition on the equilibrium
density
operator $\hat\rho$ the following relation that the
OLM approach
inherits~\cite{TMP,OLM,ley0,virial,gasideal,temperature,cuerponegro},
namely,
\begin{equation}
\tr\,\left[\hat{f}_q^{\; 1/(1-q)}\right] =
\tr\,\left[\hat{f}_q^{\;
q/(1-q)}\right], \label{relac1}
\end{equation}
which allows one to cast Tsallis' entropy, after one
has processed
it according to our constrained variational treatment,
in the
fashion
\begin{equation} \label{S2}
(a):\,\,\,S_{q} = \lnq\left(\bar{Z}_q\right)\,\,\,{\rm
and}\,\,\,(b):\,\,\,\,\,\, dS_{q} =
d[\lnq\left(\bar{Z}_q\right)].
\end{equation}

For the sake of completeness, we can write down the
generalized mean value of a
quantum operator $\hat O$ in terms of the quantal
configurational
characteristic as \be \langle\hat{O} \rangle_q =
\frac{\tr\,\left[{\hat
f_q}^{\; q/(1-q)} \, \hat{O}\right]} {\tr\,\left[{\hat
f_q}^{\;
q/(1-q)}\right]}. \label{generf} \ee

\section{The first law of thermodynamics}

We will now revisit the first law of thermodynamics from first
principles using the OLM-Tsallis formalism. We have already
presented some preliminary considerations in \cite{ley0}, looking
for the proper form of the Clausius equation in a NET context, but
assuming that the first law remained valid in such a case. This
last assumption is reasonable due to the fact that this law is
nothing but energy conservation. Anyway, the process developed in
\cite{ley0} can clearly be improved upon, as we will demonstrate
below. Another type of (related) analysis was performed by Wang
\cite{wang} using the canonical approach within
 the Curado-Tsallis formalism's strictures
\cite{ct_jpa24} (see also the Appendix), which
are now considered rather outmoded. Indeed,  the CT
formalism has
been disavowed even by its authors. In  \cite{wang} a
dependence
of the Hamiltonian with respect of  external
``displacements" is
also to be introduced in order to achieve the expected
results.
This is not the case here.

The traditional  Statistical Mechanics' treatment
of  thermodynamic's first law,
 within the canonical ensemble formulation,
assumes a dependence
of the internal energy upon both the density operator
and the
Hamiltonian of the system
(see Ref. \cite{katz}). In such a
formulation, variation
with respect to the system's Hamiltonian becomes then
mandatory in
dealing with the work term.

In this work, however, \begin{enumerate} \item using Occam's
razor, we will assume that  the internal  energy is a functional
of {\it just the density operator}. \item Additionally, we
consider a quite general ensemble, not merely   Gibb's canonical
one. \end{enumerate} We will show then that the first law is
recovered without any extra consideration. It is interesting to
notice that, as far as these authors know, this is the first time
in which the $q$-formulation leads to a thermodynamic result in a
rather cleaner (in Occam's terms) way than that of the traditional
$q=1$-treatment.

The basic ingredient needed for our purpose is
the definition of internal energy (Cf. (\ref{gener}))

\be
U_q=\frac{Tr\left(\hat{\rho}^q \hat
H\right)}{Tr\left(\hat{\rho}^q
\right)}. \label{Uq}\ee

We obtain $dU_q$ by thinking of $U_q$ (Cf. Eq.
(\ref{Uq})) as a functional of the
density operator alone and performing the
corresponding variations
\be
dU_q =
\delta_{\rho}[U_q]\, \delta [\hat\rho]=
\delta_{\rho}\left[
\frac{Tr\,(\hat{\rho}^q \hat{H})}{Tr (\hat{\rho}^q)
}\right] \delta\, \hat\rho  =
q\frac{Tr\,\left( \hat{\rho}^{q-1} (\hat H-U_q)
\delta\,[\hat\rho]\right)}{Tr\left(\hat{\rho}^q \right)},
\label{dU}
\ee
where
$\delta$ represents a variation and $\delta_{\rho}$
means variation with
respect to the density operator $\hat\rho$.
It is clear that the previous expression is a
particular case of the
evolution of any mean value
$\left<\widehat{O}_i\right>_q$ with respect
to the density operator (see Eq. (\ref{gener})). In
 general one has

\be
d\left<\widehat{O}_i\right>_q = q\frac{Tr\,\left[
\hat{\rho}^{q-1}
\left(\widehat{O}_i-\left<\widehat{O}_i\right>_q\right)
\delta\,[\hat\rho]\right]}{Tr\left(\hat{\rho}^q \right)}.
\label{dOj}
\ee

Now, from the form of $\hat\rho$ given by Eqs.
(\ref{rho}), and using
(\ref{charact}) and (\ref{Zqbar}), we are allowed to
write

\begin{equation}
\hat{\rho} = \frac{\left[{\hat{\openone} - (1-q)
\sum_{j=1}^M \, \lambda_j \;
\left(\widehat{O}_i-\left<\widehat{O}_i\right>_q\right)}\right]^{\;
1/(1-q)}}{\bar{Z}_q}.
\label{rhoexp}
\end{equation}
It is now easy to see that, because of i)

\be  \hat
H-U_q=\frac{1}{(1-q)\beta}\left(\hat{\openone}-\hat{\rho}^{1-q}\bar{Z}_q^{1-q}\right)-
\sum_{j=2}^M \, \frac{\lambda_j}{\beta} \;
\left(\widehat{O}_i-\left<\widehat{O}_i\right>_q\right),
\ee   and  ii)
Eq. (\ref{dOj}),  we can cast Eq. (\ref{dU}) in the
fashion

\be dU_q =\frac{q}{(1-q)\beta}\left(\frac{Tr\,
\hat{\rho}^{q-1}\,\delta\,\hat\rho_q}{Tr\,
\hat{\rho}^{q}}-
\bar{Z}_q^{1-q}\frac{Tr\, \delta\,\hat\rho_q}{Tr\,
\hat{\rho}^{q}}\right)-\sum_{j=2}^M \,
\frac{\lambda_j}{\beta}d
\left<\widehat{O}_i\right>_q.
 \label{der2}
\ee

Since \cite{OLM}

\be Tr\, \hat{\rho}^{q}=\bar{Z}_q^{1-q}, \label{ident}
\ee the second term
inside the brackets of Eq. (\ref{der2}) reduces itself
to $Tr\,
\delta\,\hat\rho$, which, on account of the
normalization condition \be Tr\,
\hat{\rho}=1, \ee vanishes identically. The first term
can be rephrased using
logarithmic derivatives  (and  Eq. (\ref{der2}))
leading to

\be d U_q =\frac{1}{(1-q)\beta}\, \delta\,\left(
\ln\left(Tr\,
\hat{\rho}^{q}\right)\right)-\sum_{j=2}^M \,
\frac{\lambda_j}{\beta}d
\left<\widehat{O}_i\right>_q, \label{der3} \ee so
that, minding
 Eq. (\ref{ident}) we finally obtain

\be
 d U_q =\frac{1}{\beta}\, d\,\left(
\ln\bar{Z}_q\right)-\sum_{j=2}^M \,
\frac{\lambda_j}{\beta}d
\left<\widehat{O}_i\right>_q.
\label{der4} \ee

 Remembering now  Eq. (\ref{der4}) we can
straightforwardly identify the ``heat"  and
 ``work"  terms of orthodox thermostatistics. If we
agree to call

\ba
d'Q_q &=&\frac{1}{\beta}\, d\,\left(
\ln\bar{Z}_q\right)\label{dQ}\\
dW &=&-\sum_{j=2}^M \, \frac{\lambda_j}{\beta}d
\left<\widehat{O}_i\right>_q,\label{dW} \ea we obtain \be
dU_q=d'Q_q+dW. \label{flaw} \ee Remember that $M=1$ corresponds to
the canonical ensemble (our a priori knowledge is restricted to
the mean value of the energy). In information theoretic terms {\it
work} entails changes in the expectation values of {\it other}
observables.

It becomes now  clear that we can re-formulate the
first law of
thermodynamics in a  non-extensive scenario and
recover
expressions that resemble the ones of the traditional,
extensive
 stage. Notice that, in the heat term,  {\it the
identification with the
entropy is lost!}  This is so because therein a {\it
natural
logarithm of the partition function} is involved, not
a $q-$logarithm, that would yield this putative
identification, since
(Cf. (\ref{S2}))

\begin{equation} \label{newS2}
(a):\,\,\,S_{q} = \lnq\left(\bar{Z}_q\right)\,\,\,{\rm
and}\,\,\,(b):\,\,\,\,\,\, dS_{q} =
d[\lnq\left(\bar{Z}_q\right)].
\end{equation}

   We can easily recover the heat-entropy  connection
by recourse to R\'enyi's
extensive information measure  \be
S_q^R=\frac{1}{(1-q)} \ln\left(Tr\,
\hat{\rho}^{q}\right), \ee and recast Eq. (\ref{dQ})
as
\be
 d'Q_q=\frac{1}{\beta}d S_q^R,
\ee in terms of what  has been called \cite{Abe1} the
{\em physical} inverse
temperature $\beta=1/T$ (see below). This ``physical"
character
is based on
 the fact that, appearances notwithstanding,  the Zero'th Law of Thermodynamics is
 strictly respected by the $q$-Thermostatistics
\cite{ley0}.

It is interesting to notice that the heat definition given by Eq.
(\ref{dQ}) does not lead to its extensive counterpart in the limit
$q\rightarrow 1$ due to the fact that it is written in terms of
the pseudo partition function $\bar{Z}_q$. On the other hand the
work term emerges in a quite clean fashion, without extra
considerations.

\subsection{Clausius Equation}

We start now with our Clausius considerations by
making  reference to Eq.
(\ref{flaw}). Let us
 restrict ourselves,  for the time being, to the heat
term alone, assuming that no work is being done. The
energy changes just on
account of heat transfer, i.e.,

\be \label{uno} dU_q = d'Q_q,\ee  where  the $d'$-notation
emphasizes the fact that the infinitesimal quantity on the right
hand side of Eq. (\ref{uno}) is NOT the thermodynamically
``relevant" one (Cf. Eq. (\ref{S2})(b)). This entails that we are
not guaranteed that there exists a putative state function $F$
such that its differential is the right hand side of Eq.
(\ref{uno}). We speak then of an {\it inexact} differential
\cite{reif} and denote it with $d'$.

Eq. (\ref{dQ}) is the OLM version of Clausius
equation. In writing it down we have
\be
d \ln\bar{Z}_q=
\frac{d'Q_q}{T} \label{heatR},
\ee
where we have used $\beta=1/T$.
Notice the presence of $\ln\bar{Z}_q$ rather than
$\ln_q\bar{Z}_q$
in Eq. (\ref{der4}) and compare (Cf.  Eq. (\ref{S2}))
with the
relation  $S_{q} = \lnq\left(\bar{Z}_q\right)$.

 According to Eq. (\ref{heatR}), in terms of the {\em
physical}
 Lagrange Multiplier $\beta$, the Tsallis
 formalism loses the {\it direct}
identification of its entropy with the heat term. This happens
because, in the concomitant MaxEnt's approach that yields $\hat
\rho$, the constraints are handled in a different manner than in
the TMP version \cite{OLM}. A
 direct
identification of $S_q$  with the heat term is
recovered if  the
(``natural") TMP Lagrange Multipliers $\beta^{TMP}$
are used
instead of $\beta$ (see below).

By recourse to the connection between $\ln\bar{Z}_q$
and Tsallis'
entropy $S_q$ \cite{OLM} we have now

\be (1-q) \ln\bar{Z}_q= \ln\left[1+(1-q)S_q^T\right],
\ee which allows us to
recast Clausius' equation, given by Eq. (\ref{heatR}),
in terms of Tsallis'
entropy, as \be \frac{d
S_q^T}{1+(1-q)S_q^T}=\frac{d'Q_q}{T}, \label{dST} \ee
or \ba d
S_q^T&=&\frac{d'Q_q}{\left(T/[1+(1-q)S_q^T]\right)}
=\frac{d'Q_q}{T_{TMP}} \nn \\
T_{TMP}&\equiv&\left(T/[1+(1-q)S_q^T]\right).
\label{dST11} \ea
 Eq. (\ref{dST}) was derived  by Abe {\em et al.}
\cite{Abe1} in what constituted the first attempt to reconciliate
the TMP-Tsallis formalism with equilibrium thermodynamics. This
result was not obtained, however, from first principles as here,
but starting from a convenient definition of the free energy.
Notice from Eq. (\ref{dST11}) that what we call $T_{TMP}$ is the
proper integrating factor for $d S_q^T$.  Eqs.
(\ref{dST})-(\ref{dST11}) were later re-derived in a very elegant
fashion by Toral \cite{toral}, appealing to the micro-canonical
ensemble. From still another vantage point, the work of Yamano is
to be highly recommended \cite{yamano}. Therein the connection
between statistical weights and thermodynamics is re-examined and
a detailed discussion of the first law is undertaken that appeals
to infinitesimal changes in the Hamiltonian.

\subsection{Quasi-stationary states?}

Some rather interesting conclusions can be drawn from
Eq. (\ref{dST}). The
first one is that {\em d$S_q^T$ is not well-defined at
this stage} (as a state
function) if we have to express it in terms of the
intensive temperature $T$.
 Looking
at things from another viewpoint, we can regard this
``defective" situation as
an indication that $\beta$ is not the natural
conjugate variable to the Tsallis
entropy. As we have just seen, if  we use  the TMP
temperature $T^{TMP}$.

\be \label{SS2}  \beta^{TMP} =\frac{1}{T^{TMP}} =
\frac{\partial
S_q^T}{\partial U_q}= \frac{\partial
\lnq\left(\bar{Z}_q\right)}{\partial U_q},
\ee we obtain Eq. (\ref{dST11}), that  we may
re-baptize as the TMP-Clausius
equation

\be d S_q^T=\frac{d'Q_q}{T^{TMP}}.
\label{clausiusST}\ee

We reiterate:  $T^{TMP}$,  not $T$, is the proper {\it
integrating factor} that
makes $S_q^T$  a state function and, as a consequence,
an exactly
differentiable quantity in the usual fashion
\cite{reif}.

\vskip 6mm

The simplest thermodynamic processes  are the
reversible ones that lead from a
state of equilibrium (SOE) (see Ref. \cite{Callen}) to
another SOE via a path
that runs through SOEs.
A reversible process of this kind is characterized by
the Clausius equation \cite{reif}, formally identical
to Eq. (\ref{clausiusST})

$$ d S_q^T=\frac{d'Q_q}{T^{TMP}}.$$

As has been stated above, notice however that the TMP treatment deals with
initial and final states
characterized by ``temperature"-Lagrange multipliers
that do not
 respect the Zero'th Law \cite{ley0,Abe1}. These are
then very peculiar states indeed.
 On the one hand, they have to be regarded as
 stationary ones from the point of view of
information theory, if
 only the expectation value of the Hamiltonian is
assumed to be known (canonical ensemble),
but, on the other one, from an intuitive,
thermodynamics vantage point, they can not be regarded
 as equilibrium states (because of the above mentioned Zero'th Law
violation).
{\it In this paper, we are specially interested in
these rather strange
situations} \cite{Abe2}. We {\it conjecture that we have encountered  here
quasi-stationary states}, so that
 we are dealing with a reversible
process between quasi-stationary states. This is in line
with the Tsallis' results mentioned in the Introduction.

\subsection{Two Clausius relations}

As stated above, two nonextensive-TMP
versions of the
Clausius equation exist. The
``pure" TMP version has already been discussed. We
pass now
to the OLM analysis of Eq. (\ref{dST}). In this case
the Zero'th Law is respected by
the pertinent Lagrange Multipliers, i.e., we
are dealing with states of equilibrium from the
thermodynamic point of view.
A reversible process between two equilibrium states
will be governed by Eq.
(\ref{heatR}):

$$d \ln\bar{Z}_q=\frac{d'Q_q}{T},$$
where $\ln\bar{Z}_q=S_q^R$ is an extensive entropy and
its conjugated temperature $T$
is intensive.

If we were confronting an
extensive irreversible process between two states of
equilibrium, we should have instead of Eq.
(\ref{heatR}) an
equation of the form

\be \label{renueva} d(\ln\; \bar{Z}_q)=\frac{d'Q_q}{T}+d_iS, \ee
with an extra term $d_iS$ added to the heat one representing the
spontaneous production of entropy. Let us once again focus
attention upon Eq. (\ref{dST})
$$\frac{d
S_q^T}{1+(1-q)S_q^T}=\frac{d'Q_q}{T}$$
It is
clear that Eqs. (\ref{heatR}) and (\ref{dST}) are two
manifestations of the
same equation. However,
by  adequately  rearranging terms we can cast Eq.
(\ref{dST}) in the fashion

\be
dS_q^{T}=\frac{d'Q_q}{T}\left[ 1\,+\,(1-q)S_q^{T}\right]=\frac{d'Q_q}{T}+d_iS^T, \label{dSTirrev} \ee
with
\be
d_iS^T=(1-q)S_q^{T}\frac{d'Q_q}{T}. \label{diST} \ee
The additional term on the right hand side of (\ref{dSTirrev}) vanishes for $q=1$. We face a
nitid nonextensive effect. Entropic changes depend not only on the amount of heat exchanged
and the temperature but also on the previous value of the entropy.
Comparing Eq. (\ref{dSTirrev}) with Eq.
(\ref{renueva}), this relation looks like the equation
for a non
reversible process, with  an
``entropy production" (a spontaneous entropy change
$d_iS^T$)  characterized
by 1) non-extensivity (either $d_iS^T>0$ for $1-q>0$
or, mutatis mutandi,
viceversa), 2) the information measure $S_q^T$, 3) the
heat flow $d'Q$, and 4)
the physical temperature. The
non-extensivity of the
entropy induces a seemingly ``irreversible" process.

We are thus faced with the following conundrum. Clausius Law retains its traditional aspect only if we use
the non-physical temperature $T^{TMP}$ as an integrating factor. If we introduce physical temperatures,
Clausius relation turns into (\ref{dSTirrev}). If the system is in thermal contact with a heat reservoir, the
pertinent temperature is $T$, not  $T^{TMP}$. The system's Tsallis' entropy then changes in the manner
 prescribed by  (\ref{dSTirrev}).

Finally, for the sake of illumination let us re-analyze an
irreversible
process from the standpoint of the ordinary,
extensive statistics,
but using the present notation. Since
R\'enyi's entropy
is extensive, we express the heat part of the first
law in terms
of this information measure. The pertinent
(reversible
\cite{reif}) basic equation is
\be dS_q^{R}=\frac{d'Q_q}{T}.
\label{dSRrev} \ee

If we were {\it indeed} confronting an actual
extensive irreversible process, we should have Eq.
(\ref{renueva}) instead of Eq. (\ref{heatR}).
Re-expressing (\ref{renueva})
in terms of Tsallis' entropy we would then get

\be dS_q^{T}=\frac{d'Q_q}{T}+d_iS+d_iS^T,
\label{dSne}\ee with $d_iS^T$ given
by Eq. (\ref{diST}). It is then apparent that, if we
could choose the
variables such that

\be q=1+\frac{ T}{ S_q^{T}}\frac{d_iS}{d'Q_q},
\label{q}\ee then the last
two terms in Eq. (\ref{dSne}) would cancel and the
remaining equation would read \be
dS_q^{T}=\frac{d'Q_q}{T}. \ee

 We see that in the case of a bona fide irreversible
process between two states of equilibrium,
 a proper choice of the variables could turn it into
 a reversible one in terms of Tsallis entropy. This an
interesting characteristic of
 the TMP-Tsallis
 formalism that has not been exploited yet.

\section{Conclusions}

Working within the strictures of non-extensive
thermostatistics, we have
re-derived the first Law of Thermodynamics from first
principles and proved
that the assumptions made by \cite{Abe1} were indeed
the correct ones. We have showed that
the non-extensive environment allows one to perform
the derivation in a general
ensemble and without the necessity of using an
explicit dependence on the Hamiltonian,
nor a posterior dependence of the Hamiltonian on the
external control variables. The
present work can also be regarding as erecting a solid
platform  for a proper
understanding (always within the non-extensive
scenario) of the Zeroth' law, as
done in \cite{ley0}, a work in which one tacitly
assumes the validity of First
Law. As far we know, this is  the first time in which,
working within a non
extensive thermostatistics framework,  heat- and
work-terms are obtained in a
natural manner without any ad-hoc consideration.

Finally, we performed a detailed analysis of Clausius
equation from the
q-thermostatistics viewpoint for both non-homogeneous
and homogeneous systems.
Summing up
\begin{enumerate}
\item
 a non-extensive reversible process can be  achieved
between off-equilibrium states.
\item an {\it extensive} reversible process is
equivalent, in some circumstances, to a {\it non
extensive} irreversible one. The pertinent, explicit
expression for the
``irreversible" term can be cast in terms of
well-defined quantities.
\item  a particular connection between the
pertinent variables of the problem can be established that allows
Tsallis' non-extensive statistics to ``regard" an extensive
irreversible process as is it were a reversible one.
\end{enumerate}

It is also to be noticed that, with reference to Eqs. (\ref{dW}),
the formalism allows one to reinterpret, in information-theoretic
terms, the meaning of heat and work, according of what typo of a
priori knowledge is available. If this is restricted to the mean
value of energy, its associated changes are called heat. If,
additionally, other expectation values are a priori known, their
changes are called work.

\section{Appendix: Normalization choices }

We will employ here, for the sake of simplicity, a {\it classical
notation}. Consider the physical quantity $O$ that in the
microstate $n$ ($n=1,\ldots,W$) adopts the value $o_n$. Let $p_n$
stand for the microscopic probability for the microstate $n$. The
expectation value of $O$ is evaluated in the literature according
to three distinct recipes, denoted here by $\langle
O\rangle^{(1)}$, $\langle O\rangle^{(2)}$ and $\langle
O\rangle^{(3)}$, and referred to henceforth as the
first~\cite{t_jsp52}, second~\cite{ct_jpa24}, and third
choice~\cite{pennini,TMP}, respectively.
\begin{enumerate}
\item
The first choice
\begin{equation}
\langle O\rangle^{(1)}=\sum_{n=1}^W p_n \, o_n,
\label{vm1}
\end{equation}
was the conventional one, used by Tsallis in his
seminal
paper~\cite{t_jsp52}.
\item
The second choice
\begin{equation}
\langle O\rangle^{(2)}=\sum_{n=1}^W p_n^{\ q} \, o_n,
\label{vm2}
\end{equation}
was regarded as the canonical one until quite
recently~\cite{ct_jpa24} and is the only one that is guaranteed to
yield, always, an analytical solution to the associated MaxEnt
variational problem~\cite{universal}. Notice, however, that the
average value of the identity operator is not equal to one.
Elaborated studies have been performed using this
``Curado--Tsallis
flavor"~\cite{turcos,u_pre60,ppp_pa234,tt_pa261}.
\item
Finally, nowadays most authors consider that the third
choice~\cite{pennini,TMP}, usually denoted as the
Tsallis--Mendes--Plastino~(TMP) one, is the most
appropriate
definition. It reads
\begin{equation}
\langle O\rangle^{(3)} =\frac{\sum_{n=1}^W p_n^{\ q}
\,
o_n}{\sum_{n'=1}^W p_{n'}^{\ q}} \equiv\langle
O\rangle_q.
\label{vm3}
\end{equation}
\end{enumerate}
As stated above, these definitions are to be employed
in order to
accommodate the available a priori information and
thus obtain the
pertinent probability distribution via Jaynes' MaxEnt
approach~\cite{jaynes,katz}, extremizing the
$q$-entropy $S_q$
subject to normalization $\left(\sum_{n=1}^W
p_n=1\right)$ and
prior knowledge of a set of $M$ nonextensive
expectation values
$\{\langle O_j\rangle^{(\nu)},\ j=1,\ldots,M\}$, with
$\nu=$1, 2,
or~3.

\end{document}